\documentclass[aps,pra,reprint,groupedaddress]{revtex4-2}

\usepackage{mathtools}
\usepackage{amssymb}
\usepackage{graphicx}

\begin{document}


\title{Four-Vector Optical Dirac Equation and Spin-Orbit Interaction of Structured Light}


\author{Longlong Feng$^{1}$}
\email{flonglong@mail.sysu.edu.cn}
\author{Qianfan Wu$^{1}$}
\affiliation{$^{1}$ School of Physics and Astronomy, Sun Yat-Sen University, Zhuhai 519082, China}


\date{\today}

\begin{abstract}
The spin-orbit interaction of light is a crucial concept for understanding the electromagnetic properties of a material and realizing the spin-controlled manipulation of optical fields. Achieving these goals requires a complete description of spin-dependent optical phenomena in the context of vector-wave mechanics. We develop an extended Dirac theory for optical fields in generic media, which was found to be akin to a non-Hermitian chiral-extension of massive fermions with anomalous magnetic momenta moving in an external pseudo-magnetic field. This similarity allows us to investigate the optical behaviors of a material by effective field theory methods and can find wide applications in metamaterials, photonic topological insulators, etc.  We demonstrate this method by studying the spin-orbit interaction of structured light in a spin-degenerate medium and inhomogeneous isotropic medium, which leads to both spin-orbital-Hall effects and spin-to-orbital angular momentum conversion. Of importance, our approach provides simple and clear physical insight into the spin-orbit interaction of light in generic media, and could potentially bridge our understanding of topological insulators between electronic and photonic systems.
\end{abstract}


\maketitle


\section{Introduction}

In classical physics, an electromagnetic wave can be described by its amplitude and phase over space-time. In some situations where the geometric optics approximation is valid, the correspondence principle allows us to make a simple analysis in terms of particle interpretation. The traditional approach is based on the Hamilton-Jacobi theory in which the ray system associated with waves is determined by a Hamilton structure induced from the dispersion relation. Thereby, the amplitude and phase of the waves are converted into canonical coordinates and momenta associated with the degrees of freedom of particles. In this sense, the variations of optical fields can be equivalently understood by the transverse displacement of light-beam trajectories and vice versa \cite{10.1103/physreva.82.063825, 10.1103/physreve.75.066609, 10.1103/physrevlett.103.100401}. 

Moreover, we have realized the correspondence between the polarization feature of an electromagnetic wave and the spin of a photon, which exhibits, beyond the correspondence principle, another essential aspect of the wave-particle duality in the quantum world.  In addition to the spin angular momentum of a photon (SAM), structured light can carry two types of orbital angular momentum (OAM) - the intrinsic orbital AM (IOAM) and extrinsic orbital AM (EOAM), associated with the helical wavefront and helical optical path respectively.  Theoretically, with the spin properties encoded in the classical Maxwell theory, incorporating with the non-integral topological phase\cite{10.1016/s0370-1573(96)00029-4}, we have appreciated a spin-orbital interaction (SOI) of light, a universal concept underlying a variety of spin-dependent optical phenomena\cite{10.1016/j.physrep.2015.06.003, 10.1038/nphoton.2015.203, 10.1038/nphoton.2015.201, 10.1103/physrevlett.126.243601}. The SOI of light manifests itself by the interplay and mutual conversion between these three types of optical AM. The coupling of SAM and EOAM leads to a family of spin-Hall effects (SHEs) of a photon - the spin-dependent displacement of a light-beam trajectory due to a Lorentz-like force from the Berry potential\cite{10.1103/physrevlett.93.083901, 10.1103/physreva.46.5199, 10.1038/nphoton.2008.229, 10.1088/1464-4258/11/9/094009, 10.1016/j.physleta.2004.10.035, 10.1103/physrevd.74.021701, 10.1103/physreva.92.043805}. It is noted that the global topological phase does not yield a classical force, but only makes transverse displacements on the intensity pattern of light fields. In addition, the SAM-IOAM coupling produces a helicity dependent optical vortex via the spin-to-orbital AM conversion\cite{10.1103/physreve.67.036618, 10.1103/physrevlett.99.073901, 10.1088/2040-8978/13/6/064001}, and the IOAM-EOAM coupling results in the orbital-Hall effect \cite{10.1088/1464-4258/8/9/008,10.1103/physrevlett.97.043901}. 

There have been various approaches proposed for studying the spin-Hall effect of light, among which the main method is based on the modified geometrodynamics of polarized light beyond the geometric optics approximation\cite{10.1103/physreva.46.5199, 10.1038/nphoton.2008.229, 10.1088/1464-4258/11/9/094009, 10.1016/j.physleta.2004.10.035, 10.1103/physrevd.74.021701}. However, the SOI phenomenon is only significant at sub-wavelength scales, and if structured light with IOAMs is taken into account, the current geometrodynamics approach is not applicable and a full treatment of the vector-wave mechanics becomes more essential. Motivated by this consideration, the aim of this work is to develop a four-vector optical Dirac equation from the classical source-free Maxwell equation in generic media, which will enable us to deal with spin-dependent spatial-temporal variations of optical fields within a unified theoretical framework. In Sec. II, based on a modified definition of the four-vector wavefunction of photons in helicity space, we present a detailed derivation of the optical Dirac equation in generic media and discuss some physical implications. We demonstrate the optical Dirac equation in two typical applications - spin-degenerate media and isotropic inhomogeneous media, in which, the Hamiltonian theories for the spin-orbital Hall effect and spin-orbital AM conversion  are given explicitly. Finally, we give concluding remarks in Sec. V. 

\section{Optical Dirac Equation in Generic Media}

In a general linear medium characterized by the real-valued tensors of permittivity ${\bf \epsilon}({\bf r}, t)$ and permeability ${\bf \mu({\bf r}, t)}$ (without including bi-anisotropic property), the electric field  ${\bf E}$ and magnetic field ${\bf B}$ are related to the auxiliary fields ${\bf D}$ and ${\bf H}$ by the local constitutive equation $ {\bf D} = {\bf \epsilon}\cdot {\bf E}$, $ {\bf B} = {\bf \mu}\cdot {\bf H}$. In terms of ${\bf D}$ and ${\bf B}$, the source free Maxwell equation takes the form, 
\begin{equation}\label{Maxwell}
\begin{array}{ll}
\displaystyle{\frac{\partial \mathbf{D}}{\partial t}}= \nabla \times (\mathbf{\mu}^{-1}\cdot\mathbf{B}),   &  \nabla \cdot \mathbf{D}=0 \\ \\
 \displaystyle{\frac{\partial \mathbf{B}}{\partial t}}= - \nabla \times (\mathbf{\epsilon}^{-1}\cdot\mathbf{D}), & \nabla \cdot \mathbf{B}=0
\end{array}
\end{equation}

We first introduce some necessary notations and relevant rules. According to the Maxwell equation (\ref{Maxwell}), we need to deal with the following kind of operation
\begin{equation}
\hat{H} \circ  = \nabla \times (\Pi \circ) 
\end{equation}
where $\Pi=\{\Pi_{ij},i,j=1,2,3\}$ is a tensor of rank 2, which can be either $\epsilon^{-1}$ or $\mu^{-1}$ for linear dispersive electromagnetic media. For simplicity, $\Pi$ is also assumed to be a spatially-homogeneous symmetric tensor. In the global Cartesian coordinate $ {\bf r} = (x,y,z)$ associated with the laboratory reference frame, the operation $\hat{H}$ can be converted to a matrix multiplication, 
\begin{equation}
\hat{H} = (\hat{\bf k}\cdot{\bf s}) \cdot \Pi
\end{equation} 
where $\hat{\bf k} = -i \nabla$ is the momentum operator, and ${\bf s}$ is the spin-1 operator given by the adjoint representation of $SO(3)$, i.e., $\{s_i\}_{jk} = -i\epsilon_{ijk}$. Taking account of the transversality condition for electromagnetic waves, we are currently using to decompose an electromagnetic vector by its transverse components (denoted by the subscript $\perp$) with respect to a given propagating direction, e.g. the $z$ axis with the basic vector ${\bf e}_z$. The transverse components can be further projected to the complex helicity-basis $\{{\bf e}_{\pm}\}$ satisfying $({\bf e}_z\cdot{\bf s}){\bf e}_{\pm} = \pm {\bf e}_{\pm}$, which is related to the `Cartesian frame' explicitly by ${\bf e}_{\pm} = ({\bf e}_x\pm i{\bf e}_y)/\sqrt{2}$. Accordingly, for a given vector ${\bf V}$, its components in the Cartesian frame are related to those in the helicity space by a unitary transformation $\hat{U}$, $(V_x, V_y, V_z)^{T} = \hat{U} (V_+, V_-, V_z)^T$ with
\begin{equation}
\hat{U}=\frac{1}{\sqrt{2}}\left(\begin{array}{ccc}
1 & 1 & 0 \\
\mathrm{i} & -\mathrm{i} & 0 \\
0 & 0 & \sqrt{2}
\end{array}\right)
\end{equation}

In the helicity basis, the curl operation is transformed to 
\begin{eqnarray}\label{curl}
\hat{U}^{-1} (\hat{\bf k}\cdot{\bf s}) \hat{U} &=& \left(\begin{array}{ccc}
\hat{k}_z & 0 & -\hat{k}_+ \\
0 & -\hat{k}_z & \hat{k}_- \\
-\hat{k}_- & \hat{k}_+ & 0 
\end{array}\right) \nonumber \\
&=& \left(\begin{array}{cc}
\hat{k}_z \sigma_3& -\sigma_3\hat{k}_{\perp}\\
 -(\sigma_3\hat{k}_{\perp})^{\dagger} & 0\\
\end{array}\right)
\end{eqnarray}
where $ \hat{k}_{\pm}=(\hat{k}_x\mp i\hat{k}_y)/\sqrt{2}$ and $\hat{k}_{\perp}=(\hat{k}_+, \hat{k}_-)^{T}$. 
For a symmetric matrix $\Pi$, it follows as 
\begin{eqnarray}\label{matrix}
\hat{U}^{-1} \Pi \hat{U} &=& \left(\begin{array}{ccc}
Q_0 & Q_{+2}& Q_{+1}\\
Q_{-2} & Q_0 & Q_{-1} \\
Q_{-1} & Q_{+1}& q_0
\end{array}\right) \nonumber \\
&=& \left(\begin{array}{cc}
Q_0 {\bf I} + {\bf Q}\cdot{\sigma}_{\perp} & {\bf q}\\
 {\bf q}^{\dagger} & q_0\\
\end{array}\right)
\end{eqnarray}
where ${\bf q} = \{Q_{+1}, Q_{-1}\}^{T}$ is the dipole momentum with the components $Q_{\pm 1}=(\Pi_{13}\mp i\Pi_{23})/\sqrt{2}$, $Q_0 \equiv \frac{1}{2}(\Pi_{11}+\Pi_{22})$ is the monopole, $q_0=\Pi_{33}$, ${\bf I}$ is the $2\times 2$ unity matrix, ${\bf Q}\cdot{\boldsymbol\sigma}_{\perp} = Q_{+2}\sigma_{+} + Q_{-2}\sigma_{-}$ is the surface term with the quadrupoles $Q_{\pm 2} = \frac{1}{2}(\Pi_{11} - \Pi_{22}) \mp i\Pi_{12}$, ${\boldsymbol\sigma}_{\perp}$ denotes for the transverse Pauli matrices$\{\sigma_i, i=1,2\}$, and $\sigma_{\pm} = \frac{1}{2}(\sigma_1 \pm i \sigma_2)$ by convention. 

Combining the Eqs.(\ref{curl}) and (\ref{matrix}) gives the transformation
\begin{eqnarray}
\hat{H} & \rightarrow &\hat{U}^{-1} \hat{H} \hat{U} = \hat{U}^{-1} (\hat{\bf k}\cdot{\bf s}) \hat{U} \hat{U}^{-1} \Pi \hat{U} \\ 
 &=& \left(\begin{array}{cc}
\sigma_3\bigl(\hat{k}_z(Q_0{\bf I} + {\bf Q}\cdot{\boldsymbol\sigma}_{\perp})  - \hat{\bf k}_{\perp} {\bf q}^{\dagger}\bigr) & \sigma_3(\hat{k}_z{\bf q} -\hat{\bf k}_{\perp}q_0)\\
 - (\sigma_3 {\hat{k}_{\perp}}^{\dagger})(Q_0{\bf I} + {\bf Q}\cdot{\boldsymbol\sigma}_{\perp}) &  - (\sigma_3 {\hat{k}_{\perp}}^{\dagger} ){\bf q} \nonumber \\
\end{array}\right)
\end{eqnarray}
Under the action of $\hat{H}$, the transverse components of ${\bf V}$ are given by 
\begin{equation}\label{Hperp}
(\hat{H} {\bf V})_{\perp} = \sigma_3\bigl[\bigl(\hat{k}_z(Q_0{\bf I} + {\bf Q}\cdot{\boldsymbol\sigma}_{\perp})  - \hat{\bf k}_{\perp} {\bf q}^{\dagger}\bigr) {\bf V}_{\perp} +(\hat{k}_z{\bf q} - \hat{\bf k}_{\perp}q_0) V_z\bigr]
\end{equation}

If ${\bf V}$ is a divergence-free vectorial field, the transversality condition $\nabla\cdot{\bf V}=0$ implies
\begin{equation}\label{Vl}
V_{z}= -\hat{k}_z^{-1}\hat{\bf k}_{\perp}^{\dagger}{\bf V}_{\perp}
\end{equation}
Substituting the longitudinal $z$ component Eq.(\ref{Vl}) into Eq.(\ref{Hperp})  
we can find the effective operator $\hat{H}_{\perp}{\bf V}_{\perp}=(\hat{H} {\bf V})_{\perp} $, a $2\times 2$ matrix operator acting on transverse two-vectors,    
\begin{equation}\label{qq}
\hat{H}_{\perp} = \sigma_3 \bigl[ H_0 + {\bf H} \cdot {\boldsymbol\sigma}_{\perp}\bigr]
\end{equation} 
where
\begin{equation}\label{HH}
\begin{array}{ll}
H_0=\hat{k}_z Q_0 + q_0\displaystyle{\frac{1}{\hat{k}_z}} \hat{k}_+ \hat{k}_-  - {\bf q} \cdot \hat{\bf k}_{\perp} \\ 
{\bf H}=\hat{k}_z {\bf Q}  + q_0\displaystyle{\frac{1}{\hat{k}_z}} \hat{\bf k}_Q - \hat{\bf k}_q 
\end{array}
\end{equation}
with
\begin{eqnarray} 
\hat{\bf k}_q \cdot {\boldsymbol\sigma}_{\perp} &=& 2\bigl[Q_{+1}\hat{k}_{+}\sigma_{+} + Q_{-1}\hat{k}_{-}\sigma_{-}\bigr] \nonumber \\
\hat{\bf k}_Q \cdot {\boldsymbol\sigma}_{\perp} &=& \hat{k}_{+}^2 \sigma_{+} + \hat{k}_{-}^2 \sigma_{-} \nonumber
\end{eqnarray}

Now, we are embarking on applying the above results to the Maxwell equations (\ref{Maxwell}). Taking the same notation rule as described above, we can write the Maxwell equation for the transverse components as
\begin{equation}\label{TransMaxwell}
\begin{array}{ll}
i\displaystyle{\frac{\partial \mathcal{D_{\perp}}}{\partial t}}= (H_0^B - {\bf H}^B\cdot {\boldsymbol\sigma}_{\perp}) i{\sigma}_3 {\mathcal B}_{\perp}\\ \\
i\displaystyle{\frac{\partial i\sigma_3 \mathcal{B}_{\perp}}{\partial t}}= (H_0^D + {\bf  H}^D\cdot {\boldsymbol\sigma}_{\perp}) {\mathcal D}_{\perp}
\end{array}
\end{equation}
where the superscripts $D$ and $B$ denote for the corresponding quantities obtained from the dielectric matrices $\epsilon^{-1}$ and $\mu^{-1}$ respectively. Alternative to the current definition of the wavefunction of photons using the Riemann-Silberstein vectors \cite{10.1016/j.physrep.2005.01.002, 10.1016/j.optcom.2005.11.071, 10.1103/physrevlett.120.243605, 10.1088/1367-2630/16/9/093008, 10.4236/ojm.2011.13008, 10.1103/physreva.98.043837}, we define two-vector wavefunctions of photons in the helicity space, 
\begin{equation}
{\bf \Psi}_{\pm} = {\mathcal D}_{\perp}\pm i\sigma_3 {\mathcal B}_{\perp}
\end{equation}
It is easy to see that ${\bf \Psi}_{\pm}$ correspond to the positive and negative energy states respectively.  Furthermore, combining them to form a four-vector wavefunction ${\bf \Psi} = ({\bf \Psi}_+, {\bf \Psi}_-)^T$ and introducing the following effective mass and momentum operators, 
\begin{equation}\label{Effmp}
\hat{m}_{\pm} = \displaystyle{\frac{1}{2}}(H_0^D \pm H_0^B), \quad \hat{\bf p}_{\pm} = \displaystyle{\frac{1}{2}} ({\bf H}^D \pm {\bf H^{B}}) 
\end{equation}
we can eventually arrive at a compact Dirac form of the Maxwell equation, 
\begin{eqnarray}\label{opdirac}
i\frac{\partial {\bf \Psi}_{\perp}}{\partial t} = \Bigl[\gamma_0 (\hat{m}_{+} + \gamma_5\hat{m}_{-}) +{\boldsymbol\gamma}_{\perp} \cdot (\hat{\bf p}_{+}+\gamma_5\hat{\bf p}_{-})\Bigr] {\bf \Psi}_{\perp} \label{DMeqn2}
\end{eqnarray}
Remarkably, Eq.(\ref{DMeqn2}) is akin to the algebraic non-Hermitian $\gamma_5$-extension of the massive Dirac equation for fermions with an anomalous magnetic momentum in an external magnetic field, which was introduced firstly by Bender et.al  
\cite{10.1016/j.physletb.2005.08.087, Alexandre_2015}.  The difference with the Dirac equation comes from the additional $\beta$ operation in the effective momentum term by $\gamma_i = \beta{\bf \alpha}_i$, which makes the optical field a non-Hermitian system because of the anti-Hermicity $\gamma_i^{\dagger} = -\gamma_i$. It has been realized that the $\gamma_5$ extension is a critical generalization to the Dirac model, and has a wide range of applications in physics. Actually, it constitutes the low energy model for the bulk states in topological insulators \cite{10.1103/revmodphys.90.015001,10.1007/978-3-642-32858-9,10.1103/physreva.100.053819}. This similarity will help us to understand photonic topological insulators from electronic systems. The other applications could be found in the chiral magnetic effect of non-Hermitian fermionic systems \cite{10.3390/sym12050761} and neutrino physics \cite{10.1007/jhep11(2015)111,10.1007/978-3-319-31356-6, 10.1088/1742-6596/873/1/012047}.  

For photonic systems, we observe that the emergence of the $\gamma_5$ extension is due to $\epsilon\neq\mu$,  namely the $\epsilon$-$\mu$ mismatch, which breaks the so called "spin-degenerate condition"\cite{10.1038/nmat3520},. Apparently, the additional effective mass $\gamma_5  \hat{m}_{-}$ corresponds to an external pseudoscalar potential. As usual, $\hat{m}_{-}$ must behave as a scalar under a Poincare transformation, except $\hat{m}_{-}$  is parity odd because of the negative parity of $\gamma_5$. As indicated by Eq.(\ref{HH}) and Eq.(\ref{Effmp}),  $\hat{m}_{-}$ is proportional to the linear momentum, and so is mostly parity odd.  In addition, the extra momentum term $\gamma_5 {\boldsymbol\gamma}\cdot {\bf p}_{-}$ is equivalent to an anomalous magnetic moment $\delta \mu_a$ in an external magnetic field ${\bf\mathcal B}$ by the interaction $\sim \delta\mu_a\sigma_3\otimes({\boldsymbol\sigma}\cdot {\mathcal B})$, or in the covariant form $ \frac{1}{2}\delta\mu_a\sigma^{\mu\nu}F_{\mu\nu}$\cite{DiracEquation}. Interestingly, this anomalous magnetic coupling actually predicts a pseudo-magnetic effect in photonics. Also, it is worth emphasizing here that the $\gamma_5$ extension offers an alternative perspective for understanding the electromagnetic properties of a material and related optical phenomena such as chirality of materials, spin-orbital Hall effect, spin-to-orbital AM conversions, etc. 

The following remarks are necessary.  The above optical Dirac equation is derived based on the assumption of real-constant symmetric dielectric tensors. Indeed, including either anti-symmetric components, e.g. associated with gyrotropic materials, or inhomogeneities, will make the dipole and quadrupole operators $\hat{\bf k}_q$ and $\hat{\bf k}_Q$ no longer in-plane transverse vectors, and the resulting longitudinal component would give rise to the chirality of materials and produce the optical activity as well. On the other hand, the properties of dispersive media characterized by permittivity and permeability tensors are usually frequency dependent. Obviously, for monochromatic electromagnetic waves, the four-vector optical Dirac equation is applicable and can be reduced to an eigenvalue problem in the spatial domain, as illustrated by the following two applications.  

\section{Spin-Degenerate Media}

In a spin-degenerate medium with equal permittivity and permeability tensors  $\epsilon =\mu$, the optical Dirac equation (\ref{opdirac}) reduces to 
\begin{equation}\label{Dirac}
i\frac{\partial}{\partial t} {\bf \Psi} = (\beta \hat{m}  + \beta {\boldsymbol \alpha} \cdot \hat{\bf p}){\bf \Psi}
\end{equation}
in which the effective mass term is
\begin{equation}\label{EM}
\hat{m} = Q_0\hat{k}_z + \frac{q_0}{2k_z}\hat{\bf k}_{\perp}^2 - {\bf q}\cdot\hat{\bf k}_{\perp}
\end{equation}
and the momentum components are
\begin{equation}\label{EP}
{\mathrm p}_{\pm}= Q_{\pm 2}\hat{k}_z  - 2Q_{\pm 1}\hat{k}_{\pm}  + \frac{q_0}{k_z}\hat{k}_{\pm}^2, \quad p_z=0
\end{equation} 
It is easy to verify that Eq.(\ref{Dirac}) describes a $\mathcal{PT}$ symmetric non-Hermitian system\cite{10.1088/0034-4885/70/6/r03, 10.1080/00018732.2021.1876991, 10.1038/nphys4323}, and would have both positive and negative energy solutions. For a photon with given negative energy, it can be regarded as a mirror photon with equal positive energy and opposite helicity. Actually, this mirror symmetry between positive and negative energy states can be understood by the following facts. First, the effective ``mass'' given by Eq.(\ref{EM}) is a pseudo scalar operator with odd parity ($\mathcal{P}$) and odd time-reversal symmetry ($\mathcal{T}$). Second, the global ``chiral'' transformation ${\bf \Psi} \rightarrow \gamma_5 {\bf \Psi}$  is equivalent to making an exchange between the positive and negative states. Taking account of $\gamma_5$ anti-commuting with $\gamma_0=\beta$ and $\gamma_i = \beta \sigma_i$, we can see that the optical Dirac equation is invariant under the $\gamma_5$ transformation combining with the ${\bf k} \rightarrow -{\bf k}$ (either $\mathcal{T}$ or $\mathcal{P}$) transformation. Meanwhile, the appearance of extra $\beta$ in the momentum term of the optical Dirac equation can be also understood, otherwise, this invariance will be broken. Furthermore, this symmetry, as such, is associated with the duality of Maxwell equations in vacuum. In fact, a simple analysis indicates that the duality transformation is equivalent to the $\gamma_5 \mathcal{T}$ transformation, and the equality $\epsilon=\mu$ is necessary for the duality symmetry.
 
Now, we solve the eigenvalue problem of Eq.(\ref{Dirac}) under the geometric approximation. On the leading order, only the first terms on the right-hand side of both Eqs.(\ref{EM}) and (\ref{EP}), $\sim O(k_z)$ are kept. Let $\omega = \lambda k_z$, and the energy-eigenvalue equation is found to satisfy $\lambda^{2} + \vert Q_{+2} \vert^2 = Q_0^2$. The real solutions require $ \vert Q_0 \vert > \vert Q_{+2} \vert$. As expected, the eigenvalue equation has two solutions with the positive and negative energies, obviously, which can be parametrized by $\lambda=Q_0\cos\theta$, $Q_{\pm 2} = Q_0\sin\theta e^{\pm i\phi}$. Specifically, for the positive energy solution $\lambda \ge 0$, $\theta\in [0,\pi/2]$. Using this parametrization, the eigenstates can have a Poincare sphere representation, 
\begin{equation}\label{ES}
\vert R\rangle=\left(\begin{array}{c}
\cos\displaystyle{\frac{\theta}{2}}\\
0 \\
0 \\
-\sin\displaystyle{\frac{\theta}{2}}e^{-i\phi}
\end{array}\right);
\vert L \rangle=\left(\begin{array}{c}
0 \\
\cos\displaystyle{\frac{\theta}{2}} \\
-\sin\displaystyle{\frac{\theta}{2}}e^{i\phi}\\
0
\end{array}\right)
\end{equation}
For general positive-energy solutions to Eq.(\ref{Dirac}), we expand by a linear superposition of the two helicity eigenstates (\ref{ES}),  
\begin{equation}\label{Expansion}
{\bf \Psi} = (\varphi_{R}\vert R\rangle + \varphi_{L}\vert L\rangle){\mathrm e}^{-i\omega t + ik_z z}
\end{equation}
Defining the two-vector ${\Phi} = (\varphi_{R}, \varphi_{L})^{T}$ and substituting Eq.(\ref{Expansion}) into Eq.(\ref{Dirac}), we are led to an optical Schr\"{o}dinger equation under the paraxial approximation,
\begin{equation}\label{Paxa}
iQ_0\frac{\partial \Phi}{\partial z} = -\frac{q_0}{2k_z}\nabla^2_{\perp} \Phi  + V({\bf r}_{\perp})\Phi
\end{equation}
with the complex potential
\begin{eqnarray}\label{PH}
V({\bf r}_{\perp}) &=&  - {\bf q} \cdot {\bf k}_{\perp} + \frac{q_0}{2\lambda_+k_z}\sigma_3\bigl[Q_{+2}\hat{k}_{-}^2  - Q_{-2}\hat{k}_{+}^2 \bigr]  \nonumber \\
&-& \frac{1}{\lambda_+} \sigma_3\bigl[Q_{+2}Q_{-1}\hat{k}_{-} - Q_{-2}Q_{+1}\hat{k}_{+}\bigr]
\end{eqnarray}
Apparently, the above equation is similar to the simplest optical model describing the paraxial propagation of light in an inhomogeneous dielectric medium with a complex refractive index $V=n_r+ i n_i$\cite{PhysRevLett.100.103904, 10.1002/lpor.200810055}, but what we are discussing here is a two-component system. In free space, the interaction potential vanishes, and thus Eq.(\ref{Paxa}) reduces to the familiar paraxial equation, which could have various family solutions of structured light with OAMs, e.g., the Laguerre-Gaussian (LG) modes in cylindrical coordinates. 

Obviously, the first term of the potential Eq.(\ref{PH}) is a dipole interaction, which, in the helicity basis, can be alternatively expressed by $i Q_1 (e^{-i\gamma}\nabla_{+} + e^{i\gamma}\nabla_{-}\bigr)$ where we have used $Q_{\pm 1} = Q_1 e^{\pm i\gamma}$. 
In the cylindrical coordinates, the complex differential operators $\nabla_{\pm}$  become
$\nabla_{\pm} = \frac{1}{\sqrt{2}}e^{\mp i\phi}(\partial_{\rho} \pm \frac{1}{\rho}L_z)$
where $L_z= -i\partial_\phi$ is the orbital angular momentum operator in the z-direction. Obviously, for the eigenstates $L_z \vert l\rangle = l \vert l \rangle$, $\nabla_{\pm}$ are functioning as ladder operators to lowers/raises the OAMs by one unit $\hslash$ per photon.  For instance, we simply consider
a light beam of the Hermite-Gauss mode $HG_{10}$, which is a superposition of two LG modes with the opposite OAMs, $l=\pm 1$, $HG_{10} = \frac{1}{\sqrt{2}}(\vert 0,+1\rangle_{LG}+\vert 0,-1\rangle)_{LG} $,  and has a well-known doughnut-like intensity profile.  Due to the dipole interaction, the induced modes involve one through lowering the OAM of one unit per photon, and consequently gives rise to the vortex-free Gaussian-like beams of $\vert 0,0\rangle_{LG}$ and $\vert 0,1\rangle_{LG}$ modes, producing a bright spot in the center. There are two other high-order modes with $l = \pm 2$,  each acquiring an extra phase factor $\gamma$ but with opposite signs, and thus making a global rotation of the intensity pattern. It is worth noting that the dipole interaction actually makes vortex-dependent shifts of light beams, an orbital-Hall effect.    

The last two terms give the anti-Hermitian spin-dependent coupling effects arising from the light-matter interaction.  The former comes from the quadrupole-OAM interaction, while the latter from the dipole-quadrupole-OAM interaction. The following concepts borrowed from the gravity theory would be helpful for understanding these high order interactions. In fact, it is noted that the gravitational field is a natural implementation of spin-degenerate materials with $\epsilon_{ij}=\mu_{ij}=g_{ij}$, where $g_{ij}$ is the spatial metric(e.g. \cite{PhysRevD.16.933, 10.1103/physrevd.46.5407}). In any metric gravity theory,  the linearized perturbation around the Minkowski background will yield the prediction of gravitational waves. It has been found that gravitational waves can have, at most, six distinct polarization modes, including two tensor-types(spin-2), two vector types(spin-1) and two scalar-types (spin-0)\cite{10.1103/physrevlett.30.884}. Similarly, under the multipole decomposition, a symmetric dielectric tensor can have monopole, dipole and quadrupole components in the helicity space, corresponding to the spin-0, spin-1 and spin-2 virtual excitation modes respectively. In this sense, the multipole-momentum - OAM interactions can be interpreted by the AM exchange between photons and various excitation modes of a material. For instance, the interaction $Q_{+2}\hat{k}_{-}^2$ describes a $\Delta l =+2$ transition of photons by absorbing a spin-2 tensor mode $Q_{+2}$, while for $Q_{+2}Q_{-1}\hat{k}_{-}$, it gives a $\Delta l=+1$ transition of photons by absorbing a tensor mode $Q_{+2}$ and meanwhile emitting a vector mode $Q_{-1}$. Finally, we notice that, although these two types of interactions are spin-dependent, the spin itself does not get involved in the AM-exchange transitions. As indicated in the Eq.(\ref{PH}), the opposite helicity states have decoupled, implying no helicity transition would occur. In parallel with the spin-EOAM coupling, these spin-dependent interactions give rise to spin-orbital-Hall effects of light alternatively. 

\section{isotropic Inhomogeneous Media}

Let us consider a structured light propagating along the $z$-axis in an isotropic inhomogeneous medium with permittivity $\epsilon(x,y)$ and permeability $\mu(x,y)$, which are invariant in the propagation direction. Repeating the same calculation procedures as we did for the spin-degenerate medium, we can reach the same form of the Schr\"{o}dinger equation as Eq.(\ref{Paxa}) by taking $q_0=Q_0=1$ but with a different potential given by
\begin{eqnarray}\label{inhom}
&V({\bf r}_{\perp}) = \displaystyle{\frac{1}{k_z}}\Big[i{\bf q}_{\perp}\cdot\hat{\bf k}_{\perp} - ({\bf q}_{\perp}\times\hat{\bf k}_{\perp})\cdot  {\bf e}_z \sigma_3  
+ i (q_{+}\hat{k}_{+}\sigma_{+}  \nonumber \\ &+q_{-}\hat{k}_{-}\sigma_{-}) + (\epsilon^{-1}-\mu^{-1})(\hat{k}^2_{+}\sigma_{+} 
+ \hat{k}^2_{-}\sigma_{-})\Bigr] 
\end{eqnarray}
in which
\begin{eqnarray}
{\bf q}_{\perp}&=&\frac{1}{2}\frac{\epsilon^{-1}\nabla \ln\vert\epsilon\vert + \mu^{-1}\nabla\ln \vert\mu\vert}{\epsilon^{-1}+\mu^{-1}} \\
q_{\pm} &=& \frac{\epsilon^{-1}\nabla_{\pm}\ln \vert\epsilon\vert - \mu^{-1}\nabla_{\pm}\ln \vert\mu\vert}{\epsilon^{-1}+\mu^{-1}}
\end{eqnarray}
The potential Eq.(\ref{inhom}) describes the interaction between light and the optical medium, which consists of the two Hermitian and two non-Hermitian contributions. On the right-hand side of Eq.(\ref{inhom}), the first term implies a $\mathcal{PT}$-symmetric non-Hermitian dipole interaction. The second term can be viewed as a spin-IOAM interaction $\sim {\bf L}\cdot{\bf S}$ because of the dipole momentum ${\bf q}_{\perp}$ characterizing the transverse displacement, thus ${\bf L} \propto {\bf q}_{\perp}\times{\bf k}_{\perp}$,  and also ${\bf S} \propto {\boldsymbol{\sigma}}$. In addition to the spin-EOAM coupling in inhomogeneous media\cite{10.1088/1464-4258/11/9/094009, 10.1038/nphoton.2008.229, 10.1016/j.physleta.2004.10.035, 10.1103/physrevd.74.021701}, we found a similar spin-Hall effect of light due to the spin-IOAM coupling as  predicted in the spin-degenerate media.  

The last two terms can only exist for the $\epsilon$-$\mu$ mismatch media, and give rise to the spin-to-orbital AM conversion. Physically, we first note that the dipole momentum caused by the inhomogeneities of the medium can be interpreted in terms of vector perturbations - the spin-1 excitation modes. In this sense, the third term is actually a dipole-spin-orbit interaction. For the interaction $q_+\hat{k}_+\sigma_+$, what the operator $\sigma_+$ does is to make photons undergo a spin-flip transition from the spin-down state ($s_z=-1$) to the spin-up state ($s_z=1$), which is accomplished by the AM transfer $\Delta l=-1$ from the light itself through the OAM lowering operation $\hat{k}_+$ and from the optical medium through the spin-1 dipole $q_+$ respectively. In the latter, the AM transfer to photons made by the dipole operation $q_+$ can be equivalently understood to have material particles acquire an extra AM in the opposite direction, i.e., in the left-hand direction, which implies that the structured light exerts an optical torque on the medium via the spin-to-orbital AM conversion. As a conjugate of $q_+\hat{k}_+\sigma_+$, $q_-\hat{k}_-\sigma_-$ gives an inverse process.  Similarly, the fourth term gives the direct AM exchange of 2$\hbar$ per photon between the SAM and OAM of light, regardless of the inhomogeneity of the optical medium.  

Moreover, both the third and fourth terms on the right-hand side of Eq.(\ref{inhom}) are from the $\gamma_5$ extension of the momentum term $\gamma_5\gamma\cdot {\bf p}_-$, which is equivalent to the spin magnetic momentum in an external magnetic field. Attributed to the typical axial property of magnetic fields, this interaction is even in $\mathcal{P}$ and odd in $\mathcal{T}$. The resulting time-reversal symmetry breaking will further lead to the optical activity of the medium.This example demonstrates again that the four-vector optical Dirac equation offers a simple and clear field-theory description of the spin-orbit interaction of light.

\section{Outlook}


In this study, by using a modified definition of the wavefunction of photons, we establish a four-vector optical Dirac equation from the Maxwell theory in generic media, which enables us to tackle spin-dependent phenomena in the context of vector-wave mechanics. The optical field in a general medium is found to be a non-Hermitian system, whose $\mathcal{PT}$ symmetry relies on the properties of the medium. In the same spirit of our approach, it is straightforward to generalize the optical Dirac equation for gyrotropic, biisotropic and bianisotropic materials, even for anisotropy and inhomogeneous materials. For helical optical paths, the optical Dirac equation can be also extended to general three-dimensional(3D) curvilinear coordinates\cite{1063-77611961030395-08$10.00, 10.1088/1464-4258/11/9/094009}, in which the Berry potential will be naturally introduced through the moving tetrad. This approach enables us to apply the conventional field-theory methods to probe the SOI phenomena of light at sub-wavelength scales. Moreover, the optical Dirac equation may deepen our understanding of the spin Hall effect of light from that in electron-systems, and can have wide applications in metamaterials, photonic topological insulators, non-Hermitian photonics etc.\cite{10.1038/nmat3520, 10.1038/s41377-020-0331-y, 10.1103/revmodphys.91.015006}.

\begin{acknowledgments}
FLL is supported by the National Key R\&D Program of China through grant 2020YFC2201400 and the Key Program of NFSC through grant 11733010 and 11333008.
\end{acknowledgments}


\providecommand{\noopsort}[1]{}

\end{document}